\documentclass[12pt]{article}
\pdfoutput = 1
\usepackage{amsmath,amssymb,bm,epsfig,afterpage}
\usepackage{mathtools}
\usepackage{cite}
\usepackage{here}
\usepackage{color}
\usepackage{tabularx}
\usepackage{simplewick} 
\usepackage{bbm}

\usepackage{comment}

\renewcommand{\thefootnote}{\fnsymbol{footnote}}
\newcommand{\vev}[1]{{\langle{#1}\rangle}}

\newcommand{\br}[2]{\text{Br}({#1}\to{#2})}

\newcommand{\abs}[1]{\left|{#1}\right|}

\newcommand{\order}[1]{\mathcal{O}\left({#1}\right)}
\newcommand{\Ykr}[2]{Y^{({#1})}_{#2}} 

\newcommand{\pr}{\prime}
\newcommand{\ppr}{{\prime\prime}}

\newcommand{\vph}{\varphi}

\newcommand{\eV}{\mathrm{eV}}
\newcommand{\keV}{\mathrm{keV}}
\newcommand{\MeV}{\mathrm{MeV}}
\newcommand{\GeV}{\mathrm{GeV}}
\newcommand{\TeV}{\mathrm{TeV}}

\newcommand{\CW}{{\mathrm{CW}}}
\newcommand{\eff}{{\mathrm{eff}}}

\newcommand{\ita}{\mathrm{Im}\,\tau}
\newcommand{\rta}{\mathrm{Re}\,\tau}

\newcommand{\ol}[1]{\overline{#1}}

\newcommand{\Wcal}{\mathcal{W}}

\newcommand{\natN}{\mathbb{N}}
\newcommand{\intZ}{\mathbb{Z}}

\newcommand{\SM}{\mathrm{SM}}

\definecolor{darkviolet}{rgb}{0.58, 0.0, 0.83}

\newcommand{\BL}{B\mathrm{-}L}
\newcommand{\PQ}{\mathrm{PQ}}
\newcommand{\MSbar}{\overline{\mathrm{MS}}}

\newcommand{\dec}{\mathrm{dec}}

\renewcommand{\emph}[1]{\textit{#1}}

\def\bal#1\eal{\begin{align}#1\end{align}}

\numberwithin{equation}{section}

\allowdisplaybreaks[3]
\newcolumntype{Y}{&gt;{\centering\arraybackslash}X} 

\usepackage[colorlinks=true, linkcolor=black, citecolor=black, urlcolor=black]{hyperref}

\begin{document}

\begin{titlepage}

\begin{flushright}
 {\tt
CTPU-PTC-24-13  
}
\end{flushright}

\vspace{1.2cm}
\begin{center}
{\Large
{\bf
Finite modular majoron
}
}
\vskip 2cm
Tae Hyun Jung$^{\;a}$~\footnote{thjung0720@gmail.com}  
and 
Junichiro Kawamura$^{\;a}$~\footnote{junkmura13@gmail.com}

\vskip 0.5cm

{\it $^a$
Center for Theoretical Physics of the Universe, Institute for Basic Science (IBS),
Daejeon 34051, Korea
}\\[3pt]

\vskip 1.5cm

\begin{abstract}
We point out that the accidental $U(1)_{B-L}$ symmetry can arise 
from a finite modular symmetry $\Gamma_N$ 
in the type-I seesaw. 
The finite modular symmetry is spontaneously broken in such a way  
that the residual $\mathbb{Z}^T_N$ discrete symmetry, 
associated with the $T$-transformation which shifts the modulus $\tau \to \tau+ 1$, 
remains unbroken. 
This discrete $\mathbb{Z}^T_N$ symmetry mimics $U(1)_{B-L}$, 
and hence the majoron appears as a pseudo Nambu-Goldstone boson of $U(1)_{B-L}$. 
Without introducing additional interactions, 
the modulus $\tau$ can be stabilized by the Coleman-Weinberg (CW) potential given by the Majorana mass terms of the right-handed neutrinos.
We study cosmological implications of the majoron, 
with particular interests in the dark matter and dark radiation, 
where the latter may alleviate the Hubble tension.
We also find that the CW potential can have a wide range of nearly exponential shape which prevents $\tau$ from overshooting, and makes the amount of dark radiation not too large.
\end{abstract}
\end{center}
\end{titlepage}

\tableofcontents
\clearpage

\renewcommand{\thefootnote}{\arabic{footnote}}
\setcounter{footnote}{0}

\section{Introduction}

The type-I see-saw mechanism is one of the most plausible explanations for the tiny neutrino masses, 
where the active neutrino masses are suppressed
by the Majorana mass of the right-handed neutrinos~\cite{Minkowski:1977sc,Mohapatra:1979ia,Gell-Mann:1979vob,Yanagida:1979as}. 
In this mechanism, $U(1)_{\BL}$  symmetry 
is broken by the Majorana mass term, 
and hence the breaking scale of $U(1)_{\BL}$ symmetry 
is related to that of the Majorana mass which should be much smaller than the Planck scale.  
Although $U(1)_{\BL}$ symmetry can be gauged such as the Pati-Salam model~\cite{Pati:1974yy}, 
it is also interesting to consider it as a spontaneously broken global symmetry, 
so that the Nambu-Goldstone (NG) boson, the so-called majoron, 
appears in the low-energy theory~\cite{Chikashige:1980ui,Gelmini:1980re}. 
As it couples only to the neutrinos, 
the lifetime of the majoron is so long that it can be the dark matter\,\cite{Rothstein:1992rh,Berezinsky:1993fm,Lattanzi:2007ux,Bazzocchi:2008fh,Gu:2010ys,Frigerio:2011in,Lattanzi:2013uza,Queiroz:2014yna,Dudas:2014bca,Wang:2016vfj,Garcia-Cely:2017oco,Abe:2020dut,Manna:2022gwn,Chun:2023eqc} 
and/or contributes to the dark radiation. 
The latter can be tested by the measurements of 
the Cosmic Microwave Background (CMB)~\cite{Planck:2018vyg} 
and Baryon Acoustic Oscillation (BAO)~\cite{DESI:2024mwx}. 
Such majoron dark matter would be detected in future experiments~\cite{Cuesta:2022mut,Kochocki:2023lhh,Sanguineti:2022gkb,P-ONE:2020ljt} 
searching for neutrino flux from the decay of the majoron dark matter.

Theoretically, the origin of the global $U(1)_{\BL}$ symmetry 
is a fundamental question of the majoron. 
The global $U(1)_{\BL}$ symmetry guarantees the lightness of the majoron, 
while it should be broken explicitly so that the majoron is not massless. 
In other words, $U(1)_{\BL}$ symmetry should be accidental, 
and the majoron should be a pseudo NG boson. 
This would be a result of quantum gravity 
which is believed not to respect any global symmetry~\cite{Giddings:1988cx,Abbott:1989jw,Coleman:1989zu,Kallosh:1995hi}. 
However, there is no clear explanation how it is broken, 
and thus the mass of the majoron can not be predicted from a theory. 
Also, it is unclear that which interactions are allowed 
by the accidental $U(1)_{\BL}$ symmetry. 
For instance, the majoron can not be the dark matter 
if it sizably couples to lighter Standard Model (SM) fermions 
by a symmetry breaking interaction.

In this work, we point out that the accidental $U(1)_{\BL}$ for the majoron 
is naturally realized in the finite modular symmetry. 
This symmetry was originally introduced to explain the flavor structure of quarks and leptons, where the essential idea is that Yukawa coupling constants of the SM fermions 
are promoted to modular forms 
which are holomorphic functions of the complex scalar $\tau$, named modulus,  
transformed under the finite modular symmetry~\cite{Feruglio:2017spp,Kobayashi:2018vbk,Penedo:2018nmg,Novichkov:2018nkm,Ding:2019xna,Liu:2019khw,Novichkov:2020eep,Liu:2020akv,Liu:2020msy}.
In fact, the Yukawa coupling constants depend on a modulus $\tau$ in such a way 
in heterotic orbifold models~\cite{Lauer:1989ax,Lauer:1990tm,Ferrara:1989qb,Baur:2019kwi,Nilles:2020nnc,Nilles:2020gvu} 
and magnetized D-brane models~\cite{Kobayashi:2017dyu,Kobayashi:2018rad,Kobayashi:2018bff,Ohki:2020bpo,Kikuchi:2020frp,Kikuchi:2020nxn,Hoshiya:2020hki}.  
The unitary part of the finite modular symmetries 
are isomorphic to the non-Abelian discrete symmetries~\cite{deAdelhartToorop:2011re},  
which have been intensively studied to explain the SM flavor structure~\cite{Altarelli:2010gt,Ishimori:2010au,Ishimori:2012zz,Hernandez:2012ra,King:2013eh,King:2014nza,Tanimoto:2015nfa,King:2017guk,Petcov:2017ggy,Feruglio:2019ybq,Kobayashi:2022moq}. 
Similarly to those non-Abelian discrete symmetry models, 
the modular symmetry is broken by the Vacuum Expectation Value (VEV) of the modulus $\tau$ 
instead of the flavons.  
If the modulus is stabilized near one of the fixed points of the modular symmetry, 
the non-Abelian discrete symmetry is not fully broken and 
there remains a residual Abelian discrete $\intZ_n$ symmetry. 
This Abelian discrete symmetry can be used for the Froggatt-Nielsen (FN)  
mechanism to explain the flavor hierarchy~\cite{Petcov:2022fjf,Petcov:2023vws,Kikuchi:2023jap, Abe:2023ilq,Kikuchi:2023cap,Feruglio:2021dte,Novichkov:2021evw,Abe:2023qmr,Abe:2023dvr,deMedeirosVarzielas:2023crv,Kikuchi:2023fpl}.

Our idea in this paper is that the accidental $U(1)_{\BL}$ symmetry 
is realized from the residual $\intZ_n$ symmetry, 
and hence the explicit $U(1)_{\BL}$ breaking effects arises 
in a way such that the $\intZ_n$ symmetry is unbroken. 
Since the majoron is a pseudo-NG boson protected by the finite modular symmetry, 
we call it as \textit{finite modular majoron}. 
We also note that the finite modular symmetry is flavor universal 
unlike the other flavor models, 
so that it realizes the global $U(1)_{\BL}$ symmetry accidentally.

Having a large VEV of the imaginary part of the modulus $\tau$ is essential 
to realize the accidental $U(1)_{\BL}$ symmetry from the residual $\intZ_n$ symmetry
and to make the seesaw mechanism work within the perturbative range.
Interestingly, it will turn out that 
the Majorana mass term itself generates the potential of the modulus 
by the Coleman-Weinberg (CW) mechanism~\footnote{
In general, the modulus can be stabilized by 
utilizing powers of modular forms~\cite{Kobayashi:2019xvz,Kobayashi:2019uyt,Abe:2023ylh}, 
three-form fluxes~\cite{Ishiguro:2020tmo,Ishiguro:2022pde} 
or the general $SL(2,\intZ)$ superpotential~\cite{Novichkov:2022wvg,Knapp-Perez:2023nty,Feruglio:2023uof,King:2023snq} 
written by the Klein $j$ function~\cite{Cvetic:1991qm}. 
Reference~\cite{Kobayashi:2023spx} studies how the radiative correction 
affects the tree-level potential. 
}.
This scenario is similar to the one proposed in Ref.~\cite{Higaki:2024jdk}, 
where the CW potential is generated from the mass term for the vector-like quarks 
in the KSVZ axion model~\cite{Kim:1979if,Shifman:1979if}~\footnote{
See Refs.~\cite{Kobayashi:2020oji,Ahn:2023iqa,Feruglio:2023uof,Penedo:2024gtb,Petcov:2024vph} 
for other solutions to the strong CP problem by the finite modular symmetry 
}. 
In the vector-like quark case, 
the accidental $U(1)_\PQ$ is realized due to the residual $\intZ_n$ symmetry, 
and the pseudo-NG boson is the QCD axion. 
We shall show that the right-handed neutrinos play the role of the vector-like quarks, 
and the modulus is stabilized near the fixed-point where the accidental $U(1)_{\BL}$ symmetry arises. In this model, we can explain the smallness of the Majorana mass term 
by the FN mechanism from the residual $\intZ_n$ symmetry, 
in the same way as the explanation for the flavor hierarchies. 
Furthermore, we can explicitly calculate the masses of the majoron $\sim \rta$ 
and the modulus $\sim \ita$.

We investigate cosmological implications of the majoron in this simple model mainly focusing 
on the dark matter and dark radiation.
The majoron has a typically long lifetime, so its coherent oscillation can be the dark matter.
Since the decay constant of the majoron resides at $\order{10^{16}}~\GeV$, 
the dark matter abundance is explained when 
either the majoron mass is $\order{10^{-17}}\,\eV$, 
or the oscillation starts during a matter-dominated era.
In addition, the relativistic majorons produced from the decay of the modulus $\sim \ita$  
contribute to the effective number of neutrino species $\Delta N_{\eff}$, 
which may alleviate the Hubble tension~\cite{Riess:2021jrx,Planck:2018vyg,DiValentino:2021izs,Freedman:2021ahq,Schoneberg:2021qvd,Kamionkowski:2022pkx,Freedman:2023jcz,Verde:2023lmm}. 
We study in detail the dynamics of the modulus to estimate $\Delta N_{\eff}$ 
and to examine if the modulus can successfully settle down to the potential minimum 
in the early Universe.

The rest of this paper is organized as follows. 
In Sec.~\ref{sec-fmod}, we selectively review the basic ingredients of finite modular symmetry which are necessary to understand our model introduced in Sec.~\ref{sec-model}.
Then, its cosmological implications are discussed in Sec.~\ref{sec-cosmo}, and section~\ref{sec-concl} is devoted to the summary.  
The details of the dynamics of the modulus are explained both analytically and numerically in Appendix~\ref{sec-dynamics}.

\section{Finite modular symmetry} 
\label{sec-fmod}
In this section, we briefly review the finite modular symmetry for the following discussions, 
see e.g. Refs.~\cite{Feruglio:2017spp,Kobayashi:2023zzc,Ding:2023htn} for the details. 
The modular group $\Gamma = SL(2,\intZ)$ is defined as 
\begin{align}
 \Gamma := \left\{\left.
\begin{pmatrix}
 a & b \\ c & d
\end{pmatrix}
\right| 
a,b,c,d \in \intZ, \quad ad-bc = 1
 \right\}. 
\end{align}
The elements of $SL(2,\intZ)$ 
are generated by the three elements, $S$, $T$ and $R$, 
\begin{align}
 S = 
\begin{pmatrix}
 0 & 1 \\ -1 & 0      
\end{pmatrix}, 
\quad 
 T = 
\begin{pmatrix}
 1 & 1 \\ 0 & 1      
\end{pmatrix}, 
\quad 
 R = 
\begin{pmatrix}
 -1 & 0 \\ 0 & -1      
\end{pmatrix}.  
\end{align}
These generators satisfy
$S^2 = R$, $R^2 = (ST)^3 = \mathbbm{1}$ and $TR = RT$. 
An element of $\Gamma$ transforms a complex variable $\tau$, with $\ita>0$, as
\begin{align}
\tau \to  \gamma \tau = \frac{a\tau+b}{c\tau+d}. 
\end{align}
For each generator, we have
\begin{align}
 \tau \xrightarrow{S} -\frac{1}{\tau}, 
\quad 
 \tau \xrightarrow{T} \tau + 1, 
\quad 
 \tau \xrightarrow{R} \tau. 
\end{align}
Since $\tau$ is not changed by $R$, 
we consider $\ol{\Gamma} := \Gamma/\intZ^R_2 = \mathrm{PSL(2,\intZ)}$,
with $\intZ^R_2$ being $\intZ_2$ symmetry generated by $R$.
The generators of $\ol{\Gamma}$ are $S$ and $T$ satisfying 
 $S^2 = (ST)^3 = \mathbbm{1}$.   
We consider the principal congruence group $\Gamma(N)$ defined as 
\begin{align}
 \Gamma(N) :=  \left\{ 
\begin{pmatrix}
 a & b \\ c & d 
\end{pmatrix}
\in \Gamma, 
\quad 
\begin{pmatrix}
 a & b \\ c & d 
\end{pmatrix}
 \equiv 
\begin{pmatrix}
1 & 0 \\ 0 & 1 
\end{pmatrix}
\quad 
\mathrm{mod}~N
\right\},
\end{align}
where $N \in \natN$ is called level.
Note that $R$ only exists in $\Gamma(N)$ when $N=1$ or $2$.
Thus, we have $\ol{\Gamma}(N) := \Gamma(N)/\intZ^R_2$ for $N = 1,2$ 
and $\ol{\Gamma}(N) := \Gamma(N)$ for $N \ge 3$.

The finite modular group is defined as $\Gamma_N := \ol{\Gamma}/\ol{\Gamma}(N)$, 
whose generators $S$ and $T$ satisfy
\begin{align}
 S^2 = (ST)^3 = T^N = \mathbbm{1},
 \label{eq-STinFMG}
\end{align}
where $\mathbbm{1}$ should be understood as $\mathbbm{1} \in \Gamma(N)$.
There are more relations for $N>5$, but we do not use them in this work. 
Equation~\eqref{eq-STinFMG} shows that there are Abelian discrete symmetries $\intZ^S_2$, $\intZ^{ST}_3$ and $\intZ^{T}_N$ whose fixed points are located at $\tau = i, e^{2\pi i/3}$ and $i\infty$, respectively.
Here the superscripts represent the corresponding generator. 
The modular form is a holomorphic function of $\tau$ 
which transforms under $\gamma \in \Gamma_N$ as 
\begin{align}
 \Ykr{k}{r}(\tau) \to  \Ykr{k}{r}(\gamma \tau) = (c\tau+d)^k \rho(r) \Ykr{k}{r}(\tau), 
\end{align}
where $0 \le k \in \intZ$ is the modular weight 
and $\rho(r)$ is the unitary representation matrix of the representation $r$.  
The finite modular group $\Gamma_N$ with $N\le 5$ 
is isomorphic to the permutation groups as 
$\Gamma_{2n} \simeq  S_{2+n}$ and $\Gamma_{2n+1} \simeq  A_{3+n}$ with $n=1,2$, 
and hence the representation matrices $\rho(r)$ 
are the ones in those permutation groups.

Hereafter, in this section, we take $N=3$ for concreteness.
Since $\Gamma_N \simeq A_4$, 
there are three singlet representations $1_t$, $t=0,1,2$~\footnote{
These are often denoted by $1,1^\pr, 1^{\ppr}$ in the literature. 
In our notation, the subscript $t$ corresponds to the charge under $\intZ^T_3$, 
and can be generalized to $N\ne 3$.  
}, 
whose representation matrices are given by 
\begin{align}
 \rho_S(1_t) = 1, \quad \rho_T(1_t) = w^t, 
\end{align}
and a triplet representation $3$ with 
\begin{align}
 \rho_S(3) = \frac{1}{3}
\begin{pmatrix}
 -1 & 2 & 2 \\ 2 & -1 & 2 \\ 2 & 2 & -1
\end{pmatrix}, 
\quad 
 \rho_T(3) = \mathrm{diag}\left(1,w,w^2\right),   
\end{align}
where $w := e^{2\pi i/3}$. 
Here, we consider the basis in which $\rho_T(3)$ is diagonal. 
There are $k+1$ independent functions for a given weight $k$.
For example, there is only one $3$ when $k=2$, and there are $1_0$, $1_1$ and $3$ when $k=4$, and so on.  
In particular, if we consider only $k\le 10$, the non-trivial singlet representation $1_1$ 
exists only at $k = 4,8,10$ and $1_2$ exists only at $k=8$.
These singlet representations will be directly used for our model building in the next section.

The modular form with the modular weight $k=2$ is explicitly given by~\cite{Liu:2019khw},   
\begin{align}
 \Ykr{2}{3}(\tau) = 
\begin{pmatrix}
 Y_1(\tau) \\ Y_2(\tau) \\Y_3(\tau)
\end{pmatrix}
= 
\begin{pmatrix}
(3e_1(\tau)+e_2(\tau))^2 \\ -6e_1(\tau)(3e_1(\tau)+e_2(\tau)) \\ -18 e_1^2(\tau) 
\end{pmatrix},  
\end{align}
where 
\begin{align}
 e_1(\tau) :=  \frac{\eta^3(3\tau)}{\eta(\tau)}, 
\quad 
 e_2(\tau) :=  \frac{\eta^3(\tau/3)}{\eta(\tau)}. 
\end{align}
Here, $\eta(\tau)$ is the Dedekind eta function, 
\begin{align}
 \eta(\tau) := q^{\frac{1}{24}}\prod_{n=1}^\infty (1-q^n), 
\end{align}
with $q := e^{2\pi i\tau}$. 
The $q$-expansions of the modular forms for $|q|\ll 1$ become
\begin{align}
 \Ykr{2}{3}(\tau) = 
\begin{pmatrix}
 1 + 12 q + 36q^2 + 12q^3 + \cdots \\ 
-6q^{1/3} \left(1+7q + 8 q^2 + 18 q^3 + \cdots \right)  \\ 
-18q^{2/3} \left(1+2q + 5q^2 + 4q^3 + \cdots \right)
\end{pmatrix}. 
\end{align}
We choose the normalization of $Y_{1,2,3}(\tau)$ 
such that these are matched to those in Ref.~\cite{Feruglio:2017spp}. 
The modular forms with higher weights can be constructed by multiplying those with $k=2$\,\cite{Liu:2019khw}.
For example, the singlet modular forms for $k=4,6$ are given by 
\begin{align}
 \Ykr{4}{1_0} = Y_1^2 + 2Y_2Y_3,
\quad 
 \Ykr{4}{1_1} = Y_3^2 + 2Y_1Y_2,
\quad 
 \Ykr{6}{1_0} = Y_1^3 +Y_2^3 +Y_3^3 - 3Y_1Y_2Y_3. 
\end{align}
The singlet modular forms for $k\ge 8$ are easily obtained 
by scalar multiplications of these. 
Note that, in the $q$-expansion, 
$\Ykr{k}{1_t} \propto q^{t/N}( 1 + \order{\abs{q}})$ 
for $\ita \gg 1$ which is close to the fixed point $\tau = i\infty$. 
The value of modular form is suppressed by $q^{t/N}$ 
which can be understood as the FN mechanism due to the residual $\intZ^T_N$ symmetry~\cite{Novichkov:2021evw}, 
where a small parameter induced by a flavon VEV is replaced by $q^{1/N}$. 
In our model, this mechanism can be applied to explain the hierarchy 
between the Majorana mass of right-handed neutrinos and the Planck scale.

\section{Finite modular majoron}
\label{sec-model}

\subsection{Model} 
Let us consider the type-I see-saw model with supersymmetry (SUSY).
We impose the finite modular symmetry $\Gamma_N$ 
where the modular field $\tau$ only couples to the right-handed neutrinos 
through the Majorana mass term.
The K\"ahler potential and superpotential are given by 
\begin{align}
\label{eq-KW} 
 K =&\ -h \log(-i\tau + i\tau^\dag) 
     + \sum_{i=1}^3 \frac{N^\dag_i N_i}{(-i\tau+i\tau^\dag)^{k/2}} 
      + K_{\mathrm{MSSM}},    
\\  
 W =&\ \frac{1}{2} \sum_{i=1}^3 \Lambda_{i} \Ykr{k}{1_{t}}(\tau) N_i N_i 
       + \sum_{i,j=1}^3 y_D^{ij} N_i L_j H_u +  W_{\mathrm{MSSM}}, 
\end{align}
where $h\in \natN$. Here, $N_i$'s are the chiral superfields 
which are identified as the right-handed neutrinos in the type-I seesaw mechanism, 
and transform under $\Gamma_N$ as 
\begin{align}
\label{eq-Ntlaw}
 N_i \to (c\tau+d)^{-k/2} \rho(1_{t}) N_i, 
\end{align}
such that the Lagrangian is invariant under $\Gamma_N$. 
The second term  in the superpotential of Eq.~\eqref{eq-KW} is the Dirac Yukawa coupling 
involving the right-handed neutrinos and the SM leptons. 
$K_{\mathrm{MSSM}}$ and $W_{\mathrm{MSSM}}$ are respectively 
the K\"ahler potential and superpotential of the chiral superfields
in the Minimal Supersymmetric Standard Model (MSSM).

For $\ita \gg 1$, the Majorana mass term becomes approximately
\begin{align}
  \Ykr{k}{1_t} N_i N_i \propto e^{(2\pi it/3) \tau_R} N_i N_i,    
\end{align}
where $\tau_R := \rta$. 
This shows how the modular symmetry provides $U(1)_{\BL}$ as an accidental symmetry since it is invariant under
\begin{align}
\frac{2\pi}{3} \tau_R \to \frac{2\pi}{3} \tau_R -2\alpha, \quad N_i \to e^{it\alpha} N_i,  
\end{align}
with $\alpha$ being a real transformation parameter.
The origin of the accidental $U(1)_{\BL}$ symmetry is the $\intZ^T_3$ symmetry under which $\tau$ and $N_i$ transform as
$\tau_R \to \tau_R + 1 \equiv \tau_R -2$ ($\mathrm{mod}~3$) and $N_i \to w^t N_i$.
So, the $\intZ^T_3$ charge $t$ is identified as the $\BL$ number. 
Since the true symmetry is $\intZ^T_3$, $U(1)_{\BL}$ is explicitly broken, 
and it is suppressed by $\abs{q} \ll 1$ which is a quantity 
respecting $\intZ^T_3$ but $U(1)_{\BL}$.   
The MSSM quarks and leptons transform in the same way as the right-handed neutrinos, 
given in Eq.~\eqref{eq-Ntlaw}. 
We assign the $\intZ^T_3$ charges and the moduar weights of those fields 
aligned with the $\BL$ number, 
so that the Yukawa couplings involving the SM fermions, including $y_D^{ij}$,   
are constant and are independent of the modulus $\tau$.
Note that, with this assignment, the discrete anomaly of $A_4 \simeq \Gamma_3$ 
is vanishing since the $\intZ_T^3$ charge $t$ is flavor universal for the three generations~\cite{Araki:2006sqx,Araki:2008ek}.
The modular weights aligned with the $\BL$ number will also manifest 
the anomaly cancellation. 
For simplicity, we assume that the scale of the Majorana mass term is flavor universal, 
i.e. $\Lambda_i =: \Lambda_N$.

\subsection{Radiative modulus stabilization}

The modulus field $\tau$ can be stabilized at $\ita \gg 1$, 
where the residual $\intZ^T_N$ symmetry is unbroken and the accidental $U(1)_{\BL}$ arises, 
by the CW potential generated by the Majorana mass term~\footnote{
We keep $N$ arbitrary in the following discussion, 
since the results are directly applicable for $N\ne 3$. 
The charge and weight assignment in the model will be different, 
especially for the anomaly cancellation. 
The concrete assignment for $N\ne 3$ is beyond the scope of this paper. 
}. 
This mechanism is proposed in Ref.~\cite{Higaki:2024jdk} 
utilizing the vector-like quarks in the KSVZ axion model. 
We point out that the role of the vector-like quarks 
can be replaced by the right-handed neutrinos in the type-I seesaw mechanism, 
and hence we do not need to extend the model.

The CW potential induced by the Majorana mass term is given by~\cite{Coleman:1973jx},  
\begin{align}
\label{eq-VCW}
 V_\CW =  \frac{3}{32\pi^2} \left[ 
 \left(m_{0}^2 + M_{N}^2(\tau)\right)^2 
 \left(\log\frac{m_{0}^2+M_{N}^2(\tau)}{\mu_*^2}-\frac{3}{2}\right)  
- 
M_{N}^2(\tau)^2  \left(\log\frac{M_{N}^2(\tau)}{\mu_*^2}-\frac{3}{2}\right) 
  \right], 
\end{align}
with 
\begin{align}
 M_{N}^2(\tau) 
:= \Lambda_{N}^2 \left(-i\tau+i\tau^\dag \right)^{k} 
         \abs{\Ykr{k}{1_{t}}(\tau)}^2,
\end{align} 
where $(-i\tau+i\tau^\dag)^{k}$ is multiplied due to the canonical normalization for the kinetic term in Eq.~\eqref{eq-KW}~\footnote{
In supergravity, the weight of modular form is shifted as $k \to k-h$ 
to be invariant under the modular transformation. 
In this case, the canonical normalization factor is also changed 
to $(-i\tau + i\tau^\dagger)^{k-h}$ and thus $M_N^2(\tau)$ 
is an invariant combination. 
Hence, the discussions do not change. 
}.
We introduced the soft SUSY breaking mass $m_0$ for the right-handed sneutrinos 
which is assumed to be independent of the modulus $\tau$ and flavor universal. 
To have a non-zero VEV at $\ita>1$
we need $m_0 < m_N$ since this minimum does not exist otherwise, as discussed later. 
The scale parameter $\mu_*$ is a specific renormalization scale in the $\MSbar$ scheme 
where the tree-level potential of $\tau$ is vanishing.
In a general choice of the renormalization scale $\mu$, 
there must be a tree-level potential associated with soft SUSY breaking 
to make the effective potential invariant under the choice of $\mu$.
Therefore, the scale of $\mu_*$ can be interpreted as the scale 
at which such tree-level potential is negligible compared with the CW potential. 
This might be a similar situation to the radiative breaking of $U(1)_{\PQ}$ symmetry~\cite{Moxhay:1984am}.

As shown in Ref.~\cite{Higaki:2024jdk}, 
the CW potential can have a global minimum at $\ita \gg 1$ for $t\ne 0$. 
The modular form is expanded by $q := e^{2\pi i\tau }$ as 
\begin{align}
 \Ykr{k}{1_t}(\tau) = q^{\frac{t}{N}}(1+ \beta q + \order{\abs{q}}),  
 \label{eq-q_expansion}
\end{align}
where the normalization of $\Ykr{k}{1_t}$ is chosen 
so that the coefficient of the leading term is unity. 
The coefficient $\beta$ is an integer depending on $t$ and $k$.
At the leading order, the CW potential can be approximated as
\begin{align}
\label{eq-VCWlo}
 V_\CW = \frac{3m_0^2 \Lambda_N^2}{16\pi^2} (2\tau_I)^k e^{-\frac{4\pi t}{N} \tau_I} 
 \left( \log\frac{\Lambda_N^2}{e \mu_*^2} + k \log (2\tau_I) 
     - \frac{4\pi t}{N}\tau_I \right) + \order{\abs{q}, m_0^4}, 
\end{align}    
where $\tau_I := \ita$. 
This potential has a minimum when
\begin{align}
\label{eq-minC}
\log\frac{\Lambda_N^2}{\mu^2_*} + k \log (2\tau_{I0}) - \frac{4\pi t}{N} \tau_{I0} = 0, 
\end{align} 
is satisfied. 
We denote $\tau_{I0} =  \ita_0 $ as the solution to this equation, and it can be expressed as
\begin{align}
 \tau_{I0} = 
   -\frac{kN}{4\pi t}\Wcal\left(-\frac{2\pi t}{kN}
          \left(\frac{\mu_*}{\Lambda_N}\right)^{2/k} \right),    
\end{align}
where the Lambert function $\Wcal$ satisfies $\Wcal(z)e^{\Wcal(z)} = z$, 
which has two values for $-e^{-1} < z < 0$, 
and we take the one which gives $\tau_{I0} > 1$~\footnote{
The Lambert function of this branch is often denoted as $\Wcal_{-1}$. 
}. 
As an example, we obtain $\tau_{I0} \simeq 4$ when $k=8$, $t=1$ and $\mu_*=\Lambda_N$, and $\tau_{I0}$  increases as $kN/t$ increases or $\mu_*/\Lambda_N$ decreases. 
The minimization condition can be read as 
\begin{align}
\label{eq-mNdef}
 \log \frac{m_N^2}{\mu^2_*} = 0,   
\quad 
 m_N^2 := \Lambda_N^2 (2\tau_{I0})^k e^{-\frac{4\pi t}{N} \tau_{I_0}}.   
\end{align}
Here, $m_N$ corresponds to the canonically normalized mass at the minimum. 
Thus the mass of the right-handed neutrino is generated at $\mu_*$. 
This simplifies the CW potential in Eq.~\eqref{eq-VCWlo} as
\begin{align}
\label{eq-potential}
 V_\CW = \frac{3m_0^2 m_N^2}{16\pi^2} \left[\left(\frac{\tau_I}{\tau_{I0}}\right)^k 
         e^{-\frac{4\pi t}{N} (\tau_I - \tau_{I0})} 
         \left\{-1 + k \log \left(\frac{\tau_I}{\tau_{I0}} \right)
         - \frac{4\pi t}{N} \left(\tau_I - \tau_{I0}\right) \right\}
        +1  
        \right],   
\end{align}
where we have added a constant $3m_0^2m_N^2/(16\pi^2)$ so that the potential energy is zero at the minimum.
 Note that this minimum does not exist for a large soft mass $m_0 > m_N$.  
 In this case, the leading $\tau$ dependence of the potential is 
 proportional to $M_N^2(\tau)$ 
 which is a monotonically decreasing function of $\tau_I$ 
 for $\tau_I > kN/(4\pi t)$ at which the maximum resides.  
 Thus, we restrict our case to $m_0 \ll m_N$.

The leading potential in Eq.~\eqref{eq-VCWlo} does not depend 
on $\tau_R := \rta$ due to the accidental $U(1)_{\BL}$ symmetry at $\tau_I \gg 1$.
The leading potential is given by 
\begin{align}
 V(\tau_R) \simeq 
          \frac{3m_0^2 m_N^2}{16\pi^2} 
          \beta^2 \abs{q}^{2}  
         \Bigl( 1 + \cos\left(
          4\pi \tau_R 
          \right)\Bigr) 
        ,     
\label{eq-VJ}
\end{align}
so its size is suppressed by $\abs{q}^2 = e^{-4\pi \tau_I} \ll 1$ 
compared with the potential for the $\tau_I$ direction~\footnote{
If $k$ and $t$ are not flavor universal and $\Lambda_i$'s are at the similar order, 
the leading term will be $\order{\abs{q} m_0^2 m_{N_2}^2}$,  
where $m_{N_2}$ is the canonically normalized mass of the second heaviest state.   
See Ref.~\cite{Higaki:2024jdk} in the case of non-universal case. 
In this work, we focus on the universal case which allows us to identify 
the modular symmetry to $U(1)_{\BL}$ and will have the majoron. 
}.  
This potential has the shift symmetry of $\tau_R \to \tau_R+1/2$\, 
whereas the original periodicity from the finite modular symmetry is $\tau_R \to \tau_R +1$. 
The smaller shift symmetry $\tau_R \to\tau_R +1/2$ in Eq.\,\eqref{eq-VJ} is an approximate result of the $q$ expansion \eqref{eq-q_expansion}, and is broken slightly when higher order corrections are included.

\subsection{Masses and decays}

Here, we discuss the masses and decays of the modulus $\tau = \tau_R + i\tau_I$ in our model. 
After canonically normalizing the kinetic term from the K\"ahler potential 
in Eq.~\eqref{eq-KW} at the minimum $\tau_I = \tau_{I0}$, 
the physical fields are defined as 
\begin{align}
\label{eq-canmin}
 \frac{J+i\vph}{\sqrt{2}} := \frac{\sqrt{h} M_p}{2\tau_{I0}} \tau. 
\end{align}
Since $J\propto {\rm Re}\tau$ can be considered as a pseudo-NG mode of the $U(1)_{\BL}$ symmetry breaking, 
we can identify $J$ as the majoron. 
As it originates from the finite modular symmetry, we call it \emph{finite modular majoron}.
Also we call $\vph$ as modulus.
From Eqs.~\eqref{eq-VJ} and~\eqref{eq-canmin}, 
we define the decay constant of majoron as
\begin{align}
 f_J =  \sqrt{\frac{h}{2}} \frac{M_p}{4\pi \tau_{I0}}
         \simeq 10^{16}~\GeV \times \sqrt{h} 
        \left(\frac{10}{\tau_{I0}}\right), 
\end{align}
so that Eq.\,\eqref{eq-VJ} is invariant under $J \to J + 2\pi f_J$.

The masses of $\vph$ and $J$ are given by
\begin{align}
 m_\vph \simeq \sqrt{\frac{3}{8\pi^2 h}} \abs{k-\frac{4\pi t}{N} \tau_{I0}}
         \frac{m_0 m_N}{M_p}, 
\quad 
 m_J \simeq \sqrt{\frac{6}{h}}\abs{\beta} \tau_{I0} e^{-2\pi \tau_{I0}} 
     \frac{m_0m_N}{M_p}.  
\end{align}
The $\vph$ mass can be estimated as 
\begin{align}
\label{eq-mPHJ}
 m_\vph \sim 80~\TeV\times \frac{1}{\sqrt{h}} 
           \abs{k-\frac{4\pi t}{N} \tau_{I0}} 
           \left(\frac{m_0}{10^{10}~\GeV}\right)
           \left(\frac{m_N}{10^{14}~\GeV}\right),  
\end{align}
and the mass ratio of the majoron to the modulus is
\begin{align}
\label{eq-mRatio}
 \frac{m_J}{m_\vph} \simeq 4\pi \abs{\beta} \tau_{I0} 
    \abs{k-\frac{4\pi t}{N}\tau_{I0}}^{-1} e^{-2\pi \tau_{I0}}
 \ll 1. 
\end{align}
This exponentially suppressed ratio $m_J/m_\vph$ 
is a distinctive property of the finite modular majoron.
For comparison, the mass originated from an explicit breaking 
by a higher-order term of the $U(1)_{\BL}$ breaking field, 
the mass is given by $f_J^{n}/M_p^{n-2}$ with $n$ 
being a (large) positive integer depending on the dimension 
of the explicit breaking operator.
In this case, the majoron mass strongly correlates with the decay constant, unlike our model.

The majoron $J$ can be the dark matter as its decay width is suppressed by $(m_\nu/M_p)^2$.
It could decay to neutrinos through the Majorana mass term if $m_J > 2m_\nu$, where the decay width is $\Gamma(J\to \nu\nu) \simeq m_\nu^2 m_J/M_p^2$.
This is small, and the corresponding lifetime is much longer than the age of the universe unless $m_J$ is greater than $10^{10}\,\GeV$\,\footnote{
The strong BBN constraints on majoron decaying into neutrinos\,\cite{Chang:1993yp,Kanzaki:2007pd,Pospelov:2010cw,Kawasaki:2017bqm,Blinov:2019gcj,Sabti:2019mhn,Chang:2024mvg} are not applicable to our case because the lifetime of majoron is much longer than the age of the Universe.
}.
As will be shown later, the mass of our majoron is smaller than $10\,\GeV$, 
so the majoron in our model is a good dark matter candidate.

The modulus dominantly decays to majorons through the K\"ahler potential, 
and the decay rate is given by~\cite{Higaki:2013lra}
\begin{align}
\label{eq-GamX}
\Gamma_\vph := \Gamma(\vph\to JJ) 
           = \frac{m_\vph^3}{16\pi h M_p^2} 
      \sim (20\,\sec)^{-1} \times\frac{1}{h}\left(\frac{m_\vph}{20\,\TeV}\right)^3.
\end{align}  
The corresponding decay temperature is
\begin{align}
\label{eq-TD}
 T_D =&\ \left(\frac{90}{\pi^2g_{*,\rho}(T_D)}\right)^{\frac{1}{4}} \sqrt{\Gamma_\vph M_p} 
     \sim 120~\keV \times \frac{1}{\sqrt{h}} 
                          \left(\frac{3.36}{g_{*,\rho}(T_D)}\right)^{\frac{1}{4}} 
                          \left(\frac{m_\vph}{10\,\TeV}\right)^{\frac{3}{2}},
\end{align}
where $g_{*,\rho}(T)$ is the effective number of relativistic degrees of freedom for energy density when photon temperature is $T$.
Therefore, the majoron abundance that is produced from $\vph \to JJ$ 
may result in a sizable $\Delta N_{\rm eff}$ before the last scattering surface.
We discuss the cosmology of the majoron and modulus in the next section.

The modulus will also decay to the SM fermions by the K\"ahler potential~\cite{Higaki:2013lra}, 
as we assign non-zero modular weights to the MSSM fields. 
The branching fraction to a pair of SM fermion is 
$\br{\vph}{\SM} \sim m_t^2/m_\vph^2 \sim \order{10^{-4}}$
for $m_\vph \sim 10~\TeV$, 
where the dominant decay mode is the top quark. 
Although the decays to quarks are subdominant 
and $T_D$ in Eq.~\eqref{eq-TD} is unchanged,
such late-time hadronic decays are severely constrained by Big Bang nucleosynthesis (BBN) analysis which becomes especially strong at $T_D\lesssim 10\,\keV$ as they can drive a photo-dissociation process of the deuterium abundance~\cite{Kawasaki:2004qu,Jedamzik:2006xz,Kawasaki:2008qe,Kawasaki:2017bqm}.
To avoid this, the mass of the modulus should be heavier than $\order{10\,\TeV}$~\footnote{
In our model, $M_{t\ol{t}}Y_\vph \sim \xi T_R B_t \sim 10^{-10}~\GeV$ 
for $\xi=10^{-4}$, $T_R = 10~\MeV$ and $B_t := \br{\vph}{tt} = 10^{-4}$ 
is below the limit of $\order{10^{-9}}~\GeV$~\cite{Kawasaki:2017bqm} for the modulus lifetime of $\order{10}$ s, where $\xi$ is defined in Eq.~\eqref{eq-xi}.
}.

\section{Cosmological implications} 
\label{sec-cosmo}
\subsection{Majoron oscillation as dark matter}

\begin{figure}[t]
 \centering
\includegraphics[width=0.6\hsize]{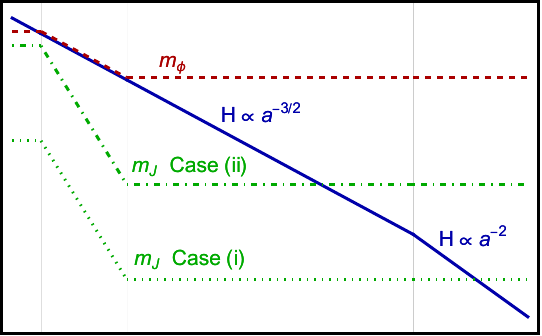}
\caption{\label{fig_sch}
Schematic picture of the evolution of the masses and the Hubble rate. 
If the majoron is ultralight ($m_J \lesssim 10^{-17}\,\eV$), 
the oscillation can start during radiation domination, 
which is the case (i), depicted by the green dot-dashed line. 
Otherwise, the oscillation should start during matter domination, 
which is the case (ii), depicted by the green dotted line. 
In any case, to avoid $\Delta N_{\rm eff}$ constraint, 
the modulus $\vph$ should start rolling down and settles down to the potential minimum 
during the matter-dominated era as depicted by the red dashed line.
The mass of $\vph$ obeys $m_\vph \sim H$ 
during the rolling along the nearly exponential potential 
before it starts to oscillate around the minimum. 
See App.~\ref{sec-dynamics} for the details. 
}
\end{figure}

As we discussed in the previous section, 
the majoron in our model is lighter than the $\GeV$ scale, 
and its lifetime is much longer than the age of the Universe.
Therefore, its coherent oscillation driven by the misalignment mechanism is a good dark matter candidate.
The large $f_J \sim \order{10^{16}}~\GeV$ strongly restricts our scenarios: 
(i) ultralight dark matter scenario with $m_J \sim 10^{-17}\,\eV$,   
or 
(ii) scenario where the oscillation starts during a matter-dominated era.
The schematic picture of the two scenarios is shown in Fig.~\ref{fig_sch}.

\subsubsection*{(i) Ultralight dark matter scenario}
If the oscillation starts during the radiation-dominated era, 
the oscillation energy density per entropy density becomes
\begin{align}
\frac{\rho_{J,{\rm osc}}}{s}
= \frac{1}{8} \left( \frac{90}{\pi^2} \right)^{\frac{1}{4}}
  \frac{g_{*,\rho}^{3/4}}{g_{*,s}} \frac{m_J^{1/2}}{M_p^{3/2}}  
   f_J^2 \theta_i^2 
 \simeq 
0.4\,\eV \times 
\left(\frac{m_J}{10^{-17}\,\eV}\right)^{1/2}
\left(\frac{f_J \theta_i}{10^{16}\,\GeV}\right)^2,
\label{eq-fuzzyDM}
\end{align}
where $\theta_i$ is the initial misalignment angle and $T_{\rm osc}$ is the temperature when the oscillation starts, i.e. $H(T_{\rm osc})=m_J$.
Here, $g_{*,s} \simeq 3.91$ 
is the effective number of relativistic degrees of freedom 
after the electron decoupling for the entropy density. 
As the observed dark matter relic corresponds to $\rho_{\rm DM}/s \simeq 0.44\,\eV$, 
our majoron can be the dark matter when the mass is around $10^{-17}\,\eV$~\cite{Planck:2018vyg}. 
Such ultralight dark matter has a de-Broglie wavelength comparable to the galactic size and is often called fuzzy dark matter\,\cite{Hu:2000ke}.
Note that, in this case, the oscillation starts at $T_{\rm osc}\sim 30\,\keV$ after the BBN.

\subsubsection*{(ii) Early matter domination scenario}

As shown in Eq.\,\eqref{eq-fuzzyDM}, 
the majoron oscillation can easily end up with the overclosure problem if $m_J>10^{-17}\,\GeV$.
To avoid this, 
we consider a matter-dominated era during which majoron starts oscillation.
The matter domination is driven by an additional field $\chi$ which we call reheaton. 
In this case, the majoron abundance is given by~\cite{Kawasaki:1995vt}
\begin{align}
\label{eq-OmegaJ}
\frac{\rho_{J,{\rm osc}}}{s} = \frac{f_J^2 \theta_i^2}{8M_p^2} T_R 
              \simeq 0.4\,\eV \times 
              \left(\frac{\theta_i}{0.14}\right)^2 
              \left(\frac{T_R}{10~\MeV}\right)
              \left(\frac{f_J}{10^{16}~\GeV}\right)^2, 
\end{align} 
so we need a mild $\order{0.1}$ tuning of the initial misalignment angle $\theta_i$.
$T_R$ is the reheating temperature after $\chi$ domination, which must be greater than $5\,\MeV$ to avoid the BBN bound as studied in Refs.\,\cite{Kawasaki:2000en,Hannestad:2004px,Ichikawa:2005vw,deSalas:2015glj,Hasegawa:2019jsa}.
Since the majoron should start oscillation before reheating, $m_J$ must be greater than $H(T_R)\simeq 4\times 10^{-14}\,\eV (T_R/10\,\MeV)^2$ for consistency.

\subsection{Modulus dynamics and dark radiation}

As the modulus $\vph$ decays to majorons around $T_D \sim \order{100\,\keV}$ 
and has the mass around the $\TeV$ scale, 
the energy density of relativistic majorons from $\vph \to JJ$ provides 
an additional relativistic degree of freedom and increases the Hubble rate at $T\lesssim T_D$. 
In the following, 
we estimate it in terms of the effective number of neutrino species $N_{\rm eff}$, 
where the estimation of the initial abundance of the modulus field is important.

In this section, we assume that $\vph$ evolves during the matter domination driven by $\chi$.
Otherwise, if the modulus starts its motion and ends up with an oscillation 
during the radiation-dominated era, 
the corresponding energy density is so large that 
it easily dominates the energy of the Universe.
In this case, the modulus domination will be finished with its decay into the majorons, 
which does not reheat the SM plasma, 
and therefore it cannot be compatible with the current Universe.
To avoid this, 
we have to restrict our scenario to the case where the modulus dynamics 
and oscillation starts during the matter-dominated era.

\begin{figure}[t]
 \centering
\begin{minipage}[t]{0.48\hsize}
\includegraphics[width=0.96\hsize]{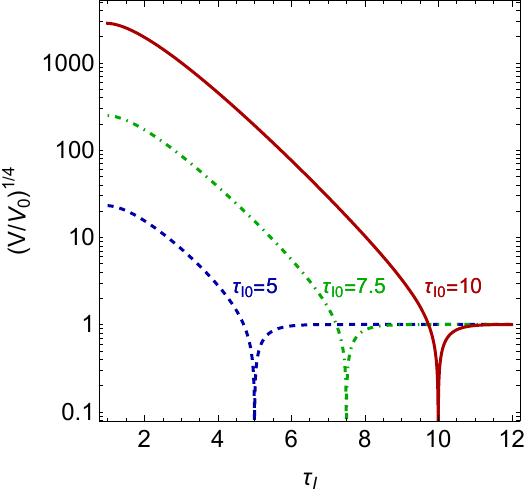}
\end{minipage}
\begin{minipage}[t]{0.48\hsize}
\includegraphics[width=0.9\hsize]{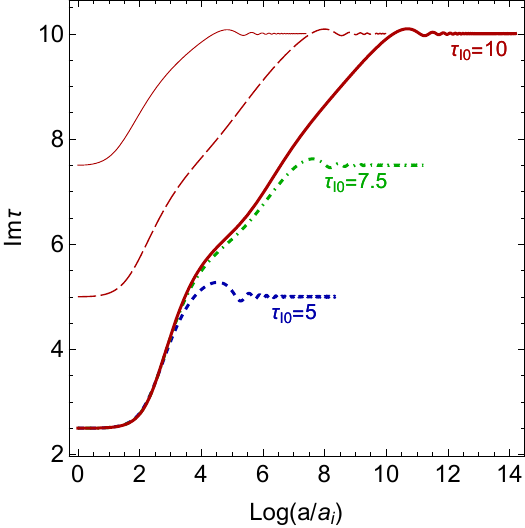}
\end{minipage}
\caption{\label{fig_VCW}
The left panel shows the shape of the potential for $t=1$, $k=4$ 
and $\tau_{I0} = 5,~7.5$ and $10$ for the blue, green and red lines, respectively.   
On the right panel, evolutions of $\ita$ from $\tau_{Ii} = 2.5$ are shown. 
The colors of the lines are the same as the left panel. 
The evolutions from $\tau_{Ii} = 5$ (dashed) and $7.5$ (thin solid) 
on the potential with $\tau_{I0} = 10$ are also shown by the red lines.  
}
\end{figure}

The modulus field $\tau_I$ (or equivalently $\vph$) has 
the nearly exponential potential $\propto \tau_I^k e^{-(4\pi t/N) \tau_I}$
which is shown in the left panel of Fig.\,\ref{fig_VCW}, 
where we depict the shape of the potential given by Eq.~\eqref{eq-potential}.
In this figure, we set the parameters as $k=4$, $h=t=1$ and $N=3$, 
and the location of the minimum is at $\tau_{I0} = 5,~7.5$ and $10$ 
on the blue, green, and red lines, respectively.   
The potential is normalized by $V_0 := 3m_0^2m_N^2/(16\pi^2)$ 
and its fourth root is plotted in the log scale. 
The potential has its maximum at 
$\tau_{I,{\rm max}} = {kN}/{(4\pi t)} \sim 1.0$, and there appears a range of $\tau_I$ between $\tau_{I,{\rm max}}$ and $\tau_{I0}$ where the potential is nearly exponential in $\tau_I$.

We solve the equation of motion numerically 
and show the evolution of $\tau_I$ as functions of $\log(a/a_i)$ 
with the scale factor $a$ in the right panel of Fig.\,\ref{fig_VCW}, 
see App.~\ref{sec-numerical} for the details of numerical calculation. 
The colors of the lines correspond to the shape of the potential 
on the left panel, i.e. the location of the minimum. 
The thick red, green, and blue lines show the evolution starting 
from $\tau_I = 2.5$, 
whereas it starts from $\tau_I = 7.5$ ($5$) for the thin red (dashed) line 
for the potential with $\tau_{I0} = 10$. 
The modulus evolves like $\tau_I \propto \log(a)$ 
and does not scale like a power of $a$ due to the Hubble friction  
when it rolls down along the exponential slope. 
Consequently, the modulus does not overshoot from the minimum 
even though its initial point is far away from the potential minimum unless the starting point is close to the maximum of the potential where its shape is no longer exponential. 

\begin{figure}[t]
 \centering
\begin{minipage}[t]{0.48\hsize}
\includegraphics[width=0.95\hsize]{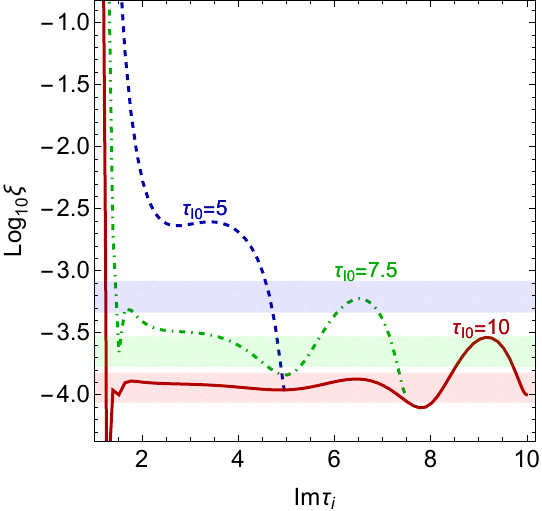}
\end{minipage}
\begin{minipage}[t]{0.48\hsize}
\includegraphics[width=0.95\hsize]{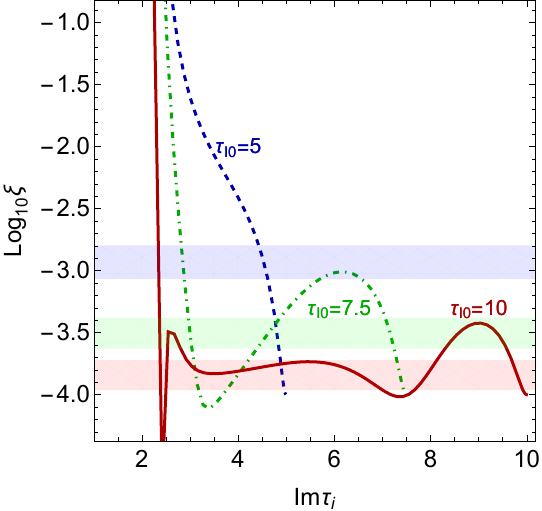}
\end{minipage}
\caption{\label{fig-vXi}
Values of $\xi$ with respect to the initial value $\tau_{Ii}$ 
for the potential with $\tau_{I0} = 5$ (blue dashed),~$7.5$ (green dotdashed) 
and~$10$ (red). 
The left panel shows the case of $k=4$ and $t=1$ 
whose potential is shown in Fig.~\ref{fig_VCW}. 
The right panels is the case of $k=8$ and $t=1$. 
The horizontal bands are the values from the analytical expression  
Eq.~\eqref{eq-xiApp} with $0.7 < F < 1.0$. 
}
\end{figure}

As shown numerically by the red line on the right panel of Fig.\,\ref{fig_VCW},  
and analytically in App.~\ref{sec-analytical}, 
the remaining oscillation after rolling down is almost independent of the initial position.
This implies an important fact that the energy density ratio between the coherent oscillation of the modulus and the SM plasma $\rho_{\rm rad}$ at the reheating temperature is given by
\begin{align}
\xi :=
\frac{\rho_{\vph,{\rm osc}} (T_R)}{\rho_{\rm rad} (T_R)}
\sim \frac{3h}{2e^2} \left(k-\frac{4\pi t}{N} \tau_{I0} \right)^{-2} 
     \sim  10^{-4} \times h\left( \frac{45}{k-4\pi t \tau_{I0}/N} \right)^2,
\label{eq-xi}
\end{align}
insensitively to its initial condition and also to the reheating temperature $T_R$.
We confirm this feature numerically as shown in Fig.~\ref{fig-vXi} which 
shows the values of $\xi$ 
with changing the initial value $\tau_{Ii} := \ita_i$. 
The left panel is in the case of the potential shown in Fig.~\ref{fig_VCW} 
with the same color coding for the location of the minimum. 
The right panel is the same figure as the left one, but $k=8$.
The colored bands are the values from our analytical estimation 
in App.~\ref{sec-analytical}.
We see that the analytical estimation nicely agrees with the numerical evaluation 
particularly for larger $\tau_{I0}$ 
in which the time of slow rolling on the exponential slope is longer. 
The value of $\xi$ does not change as long as the initial value 
is on the exponential slope. 
For smaller $\tau_{Ii}$ where it is closer to the maximum, 
where $\tau_I \sim 1.0~(1.9)$ for $k=4~(8)$, 
the modulus overshoots from the minimum and does not oscillate around the minimum. 
For $\tau_{Ii} \sim \tau_{I0}$, it is simply the oscillation around the minimum, 
and the ratio $\xi$ is not changed from the initial value, 
which is $10^{-4}$ in the plot.
In both panels, the region around the maximum, where the modulus overshoots,  
overlaps with that around the minimum, where the modulus oscillates, 
and hence there is no exponential slope for $\tau_{I0}=5$.

Now, let us estimate the modification of the Hubble rate in terms of the effective number of neutrino species.
Since, at $T<T_R$, we have the scaling behavior of $\rho_{\vph, {\rm osc}} \propto a^{-3} \propto T^3$ and $\rho_{\rm rad} \propto T^4$, their ratio scales as 
\begin{align}
\frac{\rho_{\vph, {\rm osc}}(T)}{\rho_{\rm rad}(T)} \simeq \xi \frac{T_R}{T}, 
\end{align}
which is valid until $\vph$ decays.
Assuming that $\vph$ decays instantaneously at $T=T_D$ and that $T_D<T_R$, 
we have $\rho_{J,\rm dec}/\rho_{\rm rad} = \xi (T_R/T_D)$, 
where $\rho_{J,\dec}$ is the energy density of the majoron produced from the modulus decay. 
Altogether, $\Delta N_{\rm eff}$ is given by
\begin{align}
\label{eq-delNeff}
\Delta N_{\rm eff} & = \frac{\rho_{J,\dec}}{\rho_{\nu_1}} 
   \simeq 7.4 \times \xi \frac{T_R}{T_D}
\simeq 
0.6
\times \sqrt{h} 
\left( \frac{\xi}{10^{-3}} \right)
\left( \frac{T_R}{10\,\MeV} \right)
\left( \frac{10\,\TeV}{m_\vph} \right)^{\frac{3}{2}},
\end{align}
where $\rho_{\nu_1}$ is the energy density of one species of neutrino, which is approximately $0.135 \, \rho_{\rm rad}$ for $T\ll m_e$.
Equation~\eqref{eq-GamX} is used to obtain the last equality. 
If $T_R<T_D$, $\Delta N_{\rm eff}$ is negligibly small.
Note that $\Delta N_{\rm eff}$ is sensitive to $m_\vph$ 
as $T_D$ is proportional to $m_\vph^{3/2}$ as show in Eq.\,\eqref{eq-GamX}. 
This fact indicates that $\Delta N_\eff$ has the lower bound of $\order{10^{-7}}$, 
because $m_\vph \sim m_0m_N/(4\pi M_p) \lesssim \order{10^{8}}~\GeV$  
so that $m_N < 10^{14}~\GeV$ for the perturbativity of the Dirac Yukawa coupling $y_D$, 
and $m_0 < m_N$ for the existence of the minimum. 
However, this lower bound of $\Delta N_{\rm eff}$ is negligibly small 
unless $T_R \gg \order{\MeV}$ and it seems challenging to test our model completely.

The produced majorons remains relativistic until today if 
\begin{align}
1 \ll \frac{p_J^0}{m_J} =&\ \frac{m_\vph}{2m_J} \frac{T_0}{T_D} 
\sim \ 10\times \sqrt{h}\left(\frac{\keV}{m_J}\right)
\left(\frac{10\,\TeV}{m_\vph}\right)^{1/2}, 
\end{align}
where $p_J^0$ is the momentum of the majoron today and 
$T_0 = 0.234~\mathrm{meV}$ is the temperature 
of the current Universe measured from the CMB. 
Hence, the produced majoron is relativistic 
if $m_J \lesssim \keV$ and $m_\vph \lesssim 10\,\TeV$.
For $m_J \gg p_{J0}$, 
it initially contributes to the dark radiation, but to the dark matter at a later time.
Its contribution to the dark matter relic today can be estimated as
\begin{align}
\label{eq-OmgJdec}
    \frac{\rho_{J,{\rm dec}}}{s} (T_0) 
   =  \frac{3\xi m_J}{2m_\vph} T_R
    \sim 10^{-4}~\eV \times 
      \left( \frac{\xi}{10^{-4}} \right)
       \left( \frac{m_J}{\MeV}\right)
       \left( \frac{10\,\TeV}{m_\vph}\right).
\end{align}
This amount 
is much smaller than the whole dark matter abundance $0.4~\eV$ 
for $m_J \ll m_\vph$.

The CMB analysis gives the upper bound on $\Delta N_\eff < 0.3$ 
at 95\% confidence level\,\cite{Planck:2018vyg}, 
which provides an upper bound on the modulus mass $m_\vph$ in our model, 
see Eq.~\eqref{eq-delNeff}. 
However, the boundary of this constraint is not robust 
in the presence of the Hubble tension\,\cite{Riess:2021jrx,Planck:2018vyg}, 
see also Refs.\,\cite{DiValentino:2021izs,Freedman:2021ahq,Schoneberg:2021qvd,Kamionkowski:2022pkx,Freedman:2023jcz,Verde:2023lmm} for reviews. 
For instance, $\Delta N_{\rm eff} \sim 0.4$ can alleviate the tension as analyzed in Refs.\,\cite{Bernal:2016gxb,Vagnozzi:2019ezj}. 
Therefore, $\Delta N_{\rm eff} \sim \order{0.1}$ may be preferable 
in the context of the Hubble tension, 
which can be alleviated by the majoron in our model.

\subsection{Mass spectrum and Hubble tension}
 
\begin{figure}[t]
 \centering
\begin{minipage}[t]{0.48\hsize}
\includegraphics[width=0.95\hsize]{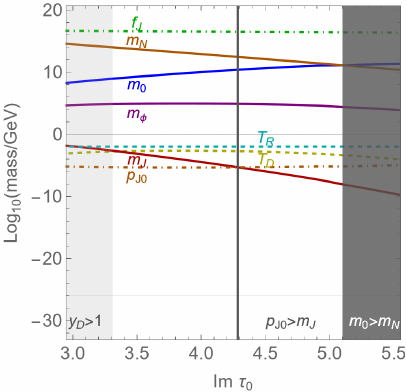}
\end{minipage}
\begin{minipage}[t]{0.48\hsize}
\includegraphics[width=0.95\hsize]{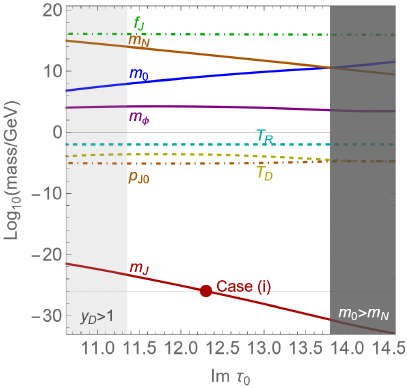}
\end{minipage}
\caption{\label{fig_spectrum}
Masses, momentum of $J$ and temperatures are depicted for $k=4$ (left) and $k=24$ (right) 
with $h=t=1$ and $\Lambda_N = M_p$.  
$m_0$ is chosen to have $\Delta N_{\mathrm{eff}} = 0.30$ when $T_R = 10~\MeV$. 
The right-handed neutrino mass is heavier than $10^{14}~\GeV$ 
on the light gray region, 
and it is lighter than the soft mass $m_0$ in the dark gray region. 
The allowed window is in between these gray regions. 
The left panel represents case (ii) where the oscillation of heavy majoron starts during early matter domination and becomes the dark matter.
The right panel corresponds to case (i) where the oscillation of ultralight majoron becomes the dark matter during radiation domination.
}
\end{figure}

\begin{figure}[t]
 \centering
\begin{minipage}[t]{0.48\hsize}
\includegraphics[width=0.95\hsize]{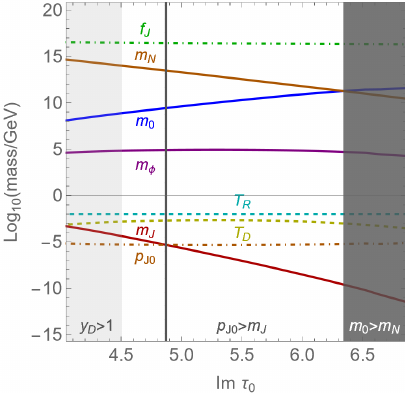}
\end{minipage}
\begin{minipage}[t]{0.48\hsize}
\includegraphics[width=0.95\hsize]{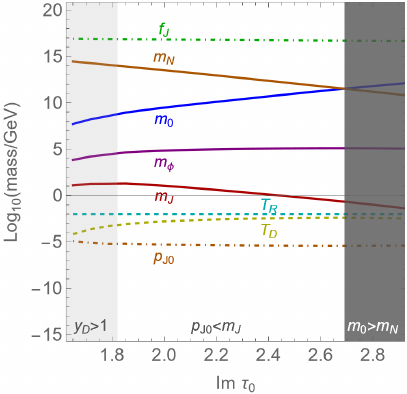}
\end{minipage}
\caption{\label{fig_spectrum8}
The same plot as Fig.~\ref{fig_spectrum}, 
but $k=8$ and $t=1$ ($t=2$) on the left (right) panel.  
}
\end{figure}

In this section, we present the mass spectrum of various particles focusing on the parameter range where $\Delta N_{\rm eff} \simeq 0.3$ is satisfied.
One can consider our choice as the $2\sigma$ boundary of the Planck constraint, or the preferred parameters that can alleviate the Hubble tension.
For example, if we take $m_0$ lower than the values presented in the following, $m_\vph$ decreases, and hence $\Delta N_{\rm eff}$ may become too large.
So, one can interpret the following values of $m_0$ and $m_\vph$ as lower bounds.
We also take the scale for the Majorana mass term as the Planck scale, i.e. $\Lambda_N = M_p$.  
The physical mass of the right-handed neutrino is given by a function of $\tau_{I0}$, and exponentially suppressed 
as $\tau_{I0}$ increases, as shown in Eq.~\eqref{eq-mNdef}. 
Similarly, $m_\vph$ is also expressed as a function of $\tau_{I0}$ and also $m_0$.
We fix $m_0$ such that $\Delta N_{\rm eff}\simeq 0.3$ by using Eqs.~\eqref{eq-GamX},~\eqref{eq-TD} and \eqref{eq-delNeff}.
Consequently, $m_\vph$ and $m_0$ are related to $\Delta N_{\eff}$ as 
\begin{align}
 m_\vph &\sim 4~\TeV\times h^{\frac{1}{3}}
                            \left(\frac{\xi}{10^{-4}}\right)^{\frac{2}{3}}
                            \left(\frac{T_R}{10~\MeV}\right)^{\frac{2}{3}} 
                    \left(\frac{0.3}{\Delta N_{\mathrm{eff}}}\right)^{\frac{2}{3}},
\\
m_0 & \sim 4\times 10^8\,\GeV
\times h^{\frac{5}{6}} \left|k-  \frac{4\pi t}{N} \tau_{I0} \right|^{-1}
\left(\frac{\xi}{10^{-4}}\right)^{\frac{2}{3}}
\left(\frac{T_R}{10\,\MeV}\right)^{\frac{2}{3}}
\left(\frac{0.3}{\Delta N_{\rm eff}}\right)^{\frac{2}{3}}
\left(\frac{10^{14}\,\GeV}{m_N}\right).
\end{align}  
The modulus mass $m_\vph$ can be greater than the $\TeV$ scale depending on $T_R$.
However, 
for the case (ii) in which the majoron oscillation starts during matter domination, 
$T_R$ should not be too far from the $10\,\MeV$ scale to avoid tuning of $\theta_i$, 
and hence $m_\vph$ should be around $\TeV$ scale.
Lastly, the majoron mass can be obtained by using Eq.~\eqref{eq-mRatio}, 
where the majoron mass is exponentially lighter than the modulus mass 
roughly by $e^{-2\pi \tau_{I0}}$.

Figure~\ref{fig_spectrum} shows the masses of the particles 
with varying the location of the minimum $\tau_{I0}$. 
The modular weight is $k=4$ ($k=24$) on the left (right) panel, 
with the other parameters are set to $\Lambda_N = M_p$, $h=t=1$, $N=3$, 
$T_R = 10~\MeV$ and $\Delta N_\eff = 0.30$. 
The initial value of $\tau_I$ is set at $\tau_{Ii} = 2.5$ ($\tau_{Ii} = 10$) 
for $k=4$ ($k=24$)~\footnote{
Regarding the modular form for $k=24$, we choose the one  
$\Ykr{24}{1_1} \propto (Y_1^2+2Y_2Y_3)^2 (Y_3^2+2Y_1Y_2)^4$ 
which gives $\beta = 448$. 
In general, the value $\beta$ is not unique for a large $k$, where there are more than one modular form for a given representation. 
}. 
In the light gray region, 
the right-handed neutrino mass is heavier than $10^{14}~\GeV$, 
and the Dirac Yukawa coupling constant for the SM neutrino $y_D$ 
is non-perturbatively large.  
Whereas, $m_N < m_0$ in the dark gray region 
where the approximation for the potential in Eq.~\eqref{eq-VCWlo} 
is broken down and the minimum disappears. 
Thus the window in between the gray regions is the viable parameter space. 
The majoron decay constant $f_J$, the majoron momentum $p_{J}^0$ 
and the temperatures $T_R$ and $T_D$ are also shown in the plot.

On the left panel where $k=4$ and $t=1$, the case (ii) can be realized, 
i.e. the majoron oscillation starts during the matter domination. 
The decay constant is always $\order{10^{16}~\GeV}$  
with slight change due to $f_J \propto \tau_{I0}^{-1}$, 
and hence $\order{0.1}$ tuning of the initial amplitude is necessary 
to explain the DM density, see Eq.~\eqref{eq-OmegaJ}.     
$T_D/T_R \sim \order{0.1\sim 0.01}$ is required to have $\Delta N_{\eff} = 0.30$, 
which can be achieved when $m_\vph$ is at $\order{\TeV}$ 
by increasing $m_0$. 
The majoron momentum is shown by the orange dot-dashed lines, 
and it is equal to the majoron mass on the vertical line at $\tau_{I0} \sim 4.9$. 
The majoron is relativistic only for the right to this line 
and contributes to $\Delta N_\eff$.

The right panel is for the case of $k=24$.  
Such a large modular weight is chosen so that 
the majoron mass is so light that 
it becomes the ultralight dark matter in case (i).
Note that the window for $m_0 < m_N < 10^{14}~\GeV$ 
shifts to larger $\tau_{I0}$ region 
since $m_N \propto \Lambda_N \tau_{I0}^{k/2} e^{-2\pi t \tau_{I0}/N}$. 
The shift of the window changes the majoron mass $\propto e^{-2\pi\tau_{I0}}$ drastically, 
whereas the other parameters are not changed significantly. 
The majorons produced from the modulus decay are always relativistic as they are ultralight.

The case of $k=8$ is also shown in Fig.~\ref{fig_spectrum8}.   
The parameters are the same as those in Fig.~\ref{fig_spectrum}, 
but $k=8$ and $t=1$ ($t=2$) on the left (right) panel. 
The initial value is $\tau_{Ii} = 3.5$ ($1.6$) for $t=1$ ($t=2$). 
We see that the mass spectrum is quite similar in the window 
except for the majoron mass,  
and
$m_N \sim 10^{10\sim 14}~\GeV$, 
$m_0 \sim 10^{8\sim 10}~\GeV$ and $m_\vph \sim 10^{0\sim 1}~\TeV$.    
Interestingly, 
the right-handed neutrino mass of the usual typical type-I seesaw is favored 
and lighter masses are not allowed to stabilize the modulus by the CW potential. 
The soft mass for the right-handed sneutrino is heavier than the observable scale, 
and hence the sparticles in the MSSM would be too heavy to be detected at the LHC, 
although it is possible that some of the sparticles are much lighter, 
such as in the split-SUSY scenario~\cite{Arkani-Hamed:2004ymt,Giudice:2004tc}.   
We note that the modulus in this scenario avoids the usual modulus problems 
due to the suppression of its energy density due to the exponential slow roll. 
In the left panel ($t=1$), 
the majorons produced from the modulus decay are relativistic 
on the right to the vertical line as in Fig.~\ref{fig_spectrum}, 
while it is always non-relativistic in the window in the right panel ($t=2$). 
Since the majoron mass is $\order{\GeV}$ in this case, 
the majoron produced from the $\vph$ decay also contributes to the DM with a sizable fraction, 
see Eqs.~\eqref{eq-OmegaJ} and~\eqref{eq-OmgJdec}. 
The existence of such exotic dark matter would give an interesting signature, but it is beyond the scope of this paper.

Before closing, we summarize the range of the majoron mass, 
which exponentially depends on the value of $\tau_{I0}$, as $e^{-2\pi \tau_{I0}}$. 
The value of $\tau_{I0}$ is determined to satisfy $m_0 < m_N < 10^{14}~\GeV$ 
where $m_N \propto \tau_{I0}^{k/2} e^{2\pi t \tau_{I0}/N}$. 
From the figures, the majoron mass is typically 
$\MeV$, $10^{-17}~\eV$, $\keV$ and $\GeV$, 
for $(k,t) = (4,1)$, $(24,1)$, $(8,1)$ and $(8,2)$, respectively. 
The majoron oscillation always contributes to the dark matter, 
while those produced from the modulus decay contribute 
to the dark matter for $m_J \gtrsim \keV$ or the dark radiation for $m_J \lesssim \keV$. 
Those two components of the majoron contribution to the dark components 
would be discriminated by the measurements of the CMB. 
If the majoron is heavy enough, 
the majoron dark matter would be probed by the searches for neutrino flux 
from the majoron decay, 
where the current limits exist for $m_J \gtrsim \order{10~\GeV}$ 
and $f_J \gtrsim \order{10^{15}~\GeV}$~\cite{Akita:2023qiz,Garcia-Cely:2017oco,KamLAND:2021gvi,Albert:2016emp,Arguelles:2022nbl}.

\section{Summary}
\label{sec-concl}
In this paper, we have proposed a simple model of finite modular majoron 
where 
the accidental $B-L$ symmetry arises from the finite modular symmetry. 
The modulus field $\tau$ only couples 
to the right-handed neutrinos in the type-I seesaw 
via the Majorana mass term being a modular form in terms of $\tau$.
If it is given by one of the non-trivial singlets under $\Gamma_N$, 
the right-handed neutrino mass is exponentially suppressed compared to the Planck scale 
as in Eq.\,\eqref{eq-mNdef}.

The accidental $U(1)_{\BL}$ symmetry is realized by the finite modular symmetry $\Gamma_N$ at the leading order of $q$-expansion, 
and hence the $\rta$ mode can be identified as the pseudo-NG boson of $U(1)_{\BL}$ symmetry breaking, which we have called majoron and denoted by $J$.  
The accidental $\BL$ symmetry could arise 
because of the residual $\intZ^T_3$ symmetry at $\ita \gg 1$, 
where the modulus transforms as $\tau \to \tau + 1$ under $\intZ^T_3$.
The majoron acquires its mass because $U(1)_{B-L}$ is an accidental symmetry 
in the leading power of $q:= e^{2\pi i\tau}$, 
and is explicitly broken in the higher order corrections at $\abs{q} \ll 1$. 
So, the majoron mass is also exponentially suppressed compared to the modulus mass despite the large decay constant at $\order{10^{16}}~\GeV$.
Since the majoron has a long lifetime, its coherent oscillation can be the dark matter.
Due to the large decay constant, there are two viable scenarios without having an overabundance; 
(i) $m_J \sim 10^{-17}\,\eV$ and becomes the ultralight dark matter,  
or 
(ii) the oscillation starts during an early matter-dominated era.

In addition, relativistic majorons 
that are produced by the decay of the modulus $\vph \propto \ita$,  
can also contribute to the dark radiation.
The dynamics of $\vph$ should start during 
an early matter-dominated era regardless of the dark matter scenarios, 
so that the amount of dark radiation is not too large. 
For the estimation of $\Delta N_{\rm eff}$, 
we have further studied the dynamics of $\vph$ both numerically and analytically 
with different initial conditions and found that 
if $\vph$ is initially located along the exponential slope of the CW potential, 
it slowly rolls down to its minimum due to the Hubble friction, 
which prevents the modulus from overshooting beyond the potential barrier.  
Another important consequence of the slow rolling is that 
the energy density of the final $\vph$ oscillation is insensitive to the initial condition, 
and is given by $\order{10^{-4}}$ of the total energy density. 
This suppression allows modulus mass heavier than $\order{\TeV}$ 
with $\Delta N_{\mathrm{eff}} = \order{0.1}$, which may alleviate the Hubble tension, 
as shown in Figs.~\ref{fig_spectrum} and~\ref{fig_spectrum8}.
Assuming the Majorana mass scale in the superpotential at the Planck scale, 
the right-handed neutrino mass $m_N$, soft right-handed sneutrino mass $m_0$ 
and the modulus mass $m_\vph$ are expected to be 
$10^{10} \lesssim m_N \lesssim 10^{14}~\GeV$, 
$10^{8} \lesssim m_0 \lesssim 10^{10}~\GeV$ 
and $m_\vph \sim \order{\TeV}$ in the phenomenologically viable parameter space.  
Whereas, the majoron mass can be 
in the range of $\keV \lesssim m_J \lesssim \GeV$ 
for the modular weight $k\le 8$.  
For a larger $k$, the majoron mass can be lighter, e.g. $m_J \sim 10^{-17}~\eV$ for $k=24$, so that it can be the ultralight dark matter.

\section*{Acknowledgement} 

J.K. thanks T.~Higaki for helpful discussions and comments on the draft.  
This work was supported by IBS under the project code, IBS-R018-D1.

\appendix

\section{Details of moduli dynamics}
\label{sec-dynamics}

The dynamics of the moduli field is important to evaluate the factor $\xi$ 
relevant to $\Delta N_{\mathrm{eff}}$. 
For the modulus whose K\"ahler potential is given by Eq.~\eqref{eq-KW},
the canonically normalized field is defined as 
\begin{align}
 \vph = \sqrt{\frac{h}{2}} M_p \log \frac{\tau_I}{\tau_{I0}}.   
\end{align} 
Note that this coincides with that in Eq.~\eqref{eq-canmin} 
at the minimum $\tau_I \sim \tau_{I0}$.
The equation of motion for $\vph$ is given by 
\begin{align}
 \ddot{\vph} +3H \dot{\vph} + \frac{\partial V(\vph)}{\partial \vph} = 0,     
\end{align}
with the Hubble parameter given by 
\begin{align}
 3M_p^2 H^2 = \frac{1}{2} \dot{\vph}^2 + V(\vph) + \rho_\chi,   
\end{align}
where $\rho_\chi$ is the reheaton energy which will be the dominant one 
in the right-hand side. 
We solve the equation of motion numerically 
as described in App.~\ref{sec-numerical}, 
and give its analytic interpretation discussed in App.~\ref{sec-analytical}.

\subsection{Numerical evaluation} 
\label{sec-numerical}

For numerical evaluation, it is convenient to introduce 
\begin{align}
 y(u) := \sqrt{\frac{2}{h}}\frac{\vph}{M_p} 
       = \log \left(\frac{\tau_I}{\tau_{I0}}\right), 
\quad 
 z(u) := \frac{H}{H_i} e^{3u} y^\pr(u), 
\end{align}
where $u:= \log(a)$ with $a$ being the scale factor,
and $\pr$ is the derivative with respect to $u$.  
$H_i$ is the initial value of the Hubble constant. 
The potential in term of $y$ is given by 
\begin{align}
 V(y) = \frac{3m_0^2 m_{N}^2}{16\pi^2} 
\left\{ 1 + e^{F(y)} (F(y)-1)\right\}, 
\end{align}
with 
\begin{align}
\label{eq-deF}
 F(y) := k y(u) - \frac{4\pi t}{N} \tau_{I0}(e^{y(u)}- 1). 
\end{align}
Here, we add the constant term for the vanishing vacuum energy at the minimum. 
The evolution equations of $y$ and $z$ are given by 
\begin{align}
y^\pr(u) =   \frac{H_i}{H} e^{-3u} z(u), 
\quad 
z^\pr(u) = - \sqrt{\frac{2}{h}} e^{3u} \frac{H_i}{H} \frac{V_\vph}{H_i^2  M_p}, 
\end{align}
where 
\begin{align}
 V_\vph(y) =&\ 
\frac{\partial V}{\partial \vph}
 =\frac{3m_0^2 m_{N}^2}{16\pi^2 \sqrt{h/2} M_p} 
\left(k- \frac{4\pi t}{N} \tau_{I0}e^y\right) 
   e^{F(y)} F(y), 
\\ \notag 
3M_p^2 H^2 =&\ 
  \frac{h}{4} H_i^2 M_p^2 e^{-6u} z^2 + V(y) + \rho_i e^{-3u}, 
\end{align}
where $\rho_i$ is the energy density of $\chi$ at $u=0$. 
The initial conditions are 
\begin{align}
 y(0) = y_i = \log\frac{\tau_{Ii}}{\tau_{I0}}, 
\quad z(0) = 0, \quad \rho_i = R \times V(y_i), 
\end{align}
with $R\gg 1$
so that we start from the universe dominated by the reheaton $\chi$. 
With this initial condition, the Hubble parameter is 
$3M_p^2 H_i^2 = (1+R) V(y_i) \simeq \rho_i$. 
We solve the equation until  $H = r m_\vph(\tau_{I0})$,  i.e. 
\begin{align}
  u = u_f := \frac{2}{3} \log \frac{H_i}{r m_\vph}, 
\end{align}
with $r \ll 1$. 
For numerical evaluation, the energy density ratio $\xi = \rho_\vph/\rho_\chi$ 
after the time of $\vph$ oscillation starts is evaluated by 
\begin{align}
\xi = \vev{\xi(u_f)} =  \frac{1}{\Delta u}\int^{u_f}_{u_f-\Delta u} du \;
   \frac{\rho_\vph(u)}{\rho_\chi(u)}, 
\end{align}
with an interval $\Delta u$. 
In our numerical analysis, $R=10^4$, $r = 10^{-3}$ and $\Delta u = 0.1$.

\subsection{Quasi-tracker solution}
\label{sec-analytical}

As the potential is nearly exponential, 
the equation of motion may follow the tracker solution~\cite{Wetterich:1987fm,Copeland:1997et,Ferreira:1997hj,Conlon:2022pnx} 
with small differences, so we call it a quasi-tracker solution.
We define the variables 
\begin{align}
 X = \frac{\dot{\vph}}{\sqrt{6}M_p H}, \quad Y = \frac{\sqrt{V}}{\sqrt{3} M_p H}, 
\end{align}
where $X^2$ and $Y^2$ correspond to the energy fraction 
of the kinetic and potential energies, respectively. 
Assuming that the reheaton energy behaves matter-like, i.e. 
\begin{align}
 \dot{\rho_\chi} + 3H \rho_\chi= 0,   
\end{align}
then the evolution equations of $X$ and $Y$ are given by 
\begin{align}
 X^\pr =&\ \frac{3}{2} \left( -1 + X^2 - Y^2 \right)   X 
            - \sqrt{\frac{3}{2}} M_p \frac{V_\vph}{V} Y^2, \\ 
Y^\pr =&\ \frac{3}{2} \left(1 + X^2 - Y^2\right)  Y 
            + \sqrt{\frac{3}{2}} M_p \frac{V_\vph}{V} X Y,  
\end{align}
where $V_\vph := \partial V/\partial \vph$.
In the case of exponential potential, 
$V\propto e^{-\lambda \vph/M_p}$ with $\lambda$ being a constant,   
the equations have the attractive tracker solution  
$X = Y = \sqrt{3/2}\lambda^{-1}$ satisfying $X^\pr = Y^\pr = 0$.

In our case, we assume that $V_\vph/V$ is slowly changing 
and the form of the tracker solution is hold. 
The solution is 
\begin{align}
\label{eq-qsol}
 X = Y = \sqrt{\frac{3}{2}}\frac{V}{M_p V_\vph} 
       = \frac{\sqrt{3h}}{2} \left(k-\frac{4\pi t}{N} \tau_I \right)^{-1} 
         \frac{e^F (F-1) + 1}{F e^F}, 
\end{align}
where $F$ is defined in Eq.~\eqref{eq-deF}. 
The potential is nearly exponential where $F \gg 1$ and is far from the minimum,  
and hence the last factor of Eq.~\eqref{eq-qsol} is almost constant.  
Thus the change of the ratio $V_\vph/V$ is only from the shift of $\tau_{I0}$ 
in the second factor  which is assumed to be slow.

For $X^2, Y^2 \ll 1$ i.e. energy domination by the reheaton,  
value of $\xi$ is estimated as 
\begin{align}
\label{eq-xiApp}
 \xi \sim  X^2 + Y^2   
   = \frac{3h}{2} \left(k-\frac{4\pi t}{N} \tau_I \right)^{-2} 
       \left( \frac{e^F(F-1)+1}{Fe^F}\right)^2,   
\end{align}
while the modulus dynamics follows the quasi-tracker solution. 
The value of $\tau_I$ in terms of $F$ is given by 
\begin{align}
 \frac{\tau_I}{\tau_{I0}}  = -\frac{kN}{4\pi t \tau_{I0}}\Wcal\left( 
  -\frac{4\pi t \tau_{I0}}{kN} \exp\left[\frac{1}{k}\left(F - \frac{4\pi t\tau_{I0}}{N}\right)
\right] \right). 
\end{align}
Then, $\xi$ does not follow the solution as it closes to the minimum 
and starts to oscillation around the minimum. 
Since the ratio $\xi = \rho_\vph/\rho_\chi$ 
is a constant during the oscillation for $T>T_R$, 
the value of $\xi$ is frozen at this time. 
We assume that the oscillation starts when $F=1$ correspond 
to the time when the potential value is same as the potential barrier. 
Finally, we estimate the $\xi$ factor after the oscillation starts as 
\begin{align}
 \xi \sim \frac{3h}{2
e^2} \left(k-\frac{4\pi t}{N} \tau_{I0} \right)^{-2} 
     \sim  10^{-4} \times \left( \frac{45}{k-4\pi t \tau_{I0}/N} \right)^2.   
\end{align}
Here, we neglect the difference between $\tau_{I0}$ and $\tau_{I}$ at $F=1$. 
Since the value of $\xi$ is determined by the value of $\tau_{I}$ 
when it starts to oscillate, it is almost independent of the initial value 
as long as it follows the quasi-tracker solution.

{\small
\bibliography{ref_modular} 
\bibliographystyle{JHEP} 
}

\end{document}